\documentclass[reprint,amsmath,amssymb,aps]{revtex4-2}
\usepackage[colorlinks,citecolor=blue,linkcolor=blue,anchorcolor=blue,filecolor=blue,
urlcolor=blue]{hyperref}
\usepackage{graphicx}
\usepackage{ulem}
\usepackage{dcolumn}
\usepackage{bm}
\usepackage{todonotes}
\newcommand{\kin}{\mathrm{kin}}
\newcommand{\dm}{\mbox{\tiny DM}}
\newcommand{\had}{\mbox{\tiny HAD}}

\begin{document}

\title{Gravitational wave asteroseismology of dark matter hadronic stars}

\author{César V. Flores}
\email{cesar.vasquez@uemasul.edu.br}
\affiliation{Universidade Estadual da Região Tocantina do Maranhão, UEMASUL, Centro de Ciências Exatas, Naturais e Tecnológicas,  Imperatriz, CEP 65901-480, MA, Brazil}
\affiliation{Universidade Federal do Maranhão, UFMA, Departamento de Física-CCET, Campus Universitário do Bacanga, São Luís, MA, CEP 65080-805, Brazil}
\author{C. H. Lenzi}
\email{chlenzi@ita.br}
\author{M. Dutra}
\email{marianad@ita.br}
\author{O. Louren\c{c}o}
\email{odilon.ita@gmail.com}
\affiliation{Departamento de F\'isica, Instituto Tecnol\'ogico de Aeron\'autica, DCTA, 12228-900, S\~ao Jos\'e dos Campos, SP, Brazil} 
\author{Jos\'e D. V. Arba\~nil}
\email{jose.arbanil@upn.pe}
\affiliation{Departamento de Ciencias, Universidad Privada del Norte, Avenida el Sol 461 San Juan de Lurigancho, 15434 Lima, Peru}
\affiliation{Facultad de Ciencias F\'isicas, Universidad Nacional Mayor de San Marcos, Avenida Venezuela s/n Cercado de Lima, 15081 Lima, Peru}

\begin{abstract}
The influence of the dark matter mass~($M_{\chi}$) and the Fermi momentum~($k_{F}^{\dm}$) on the $f_0$-mode oscillation frequency, damping time parameter, and tidal deformability of hadronic stars are studied by employing a numerical integration of hydrostatic equilibrium, nonradial oscillation, and tidal deformability equations. The matter inside the hadronic stars follows the NL3* equation of state. We obtain that the influence of $M_{\chi}$ and $k_F^{\dm}$ is observed in the $f_0$-mode, damping tome parameter, and tidal deformability. Finally, the correlation between the tidal deformability of the GW$170817$ event with $M_{\chi}$ and $k_F^{\dm}$ are also investigated.
\end{abstract}


 \maketitle

\section{Introduction}

In astrophysics, it is estimated that dark matter (DM) corresponds to approximately $27\%$ of the matter in the universe today. The name coined for this matter refers to the fact that it appears not to interact with light or electromagnetic field, hence the difficulty in its direct detection. However, this matter would interact gravitationally, thus leaving unequivocal observational signals of its existence \cite{planck}.

The study of dark matter began in the $1930$s when Fritz Zwicky observed that the velocities dispersion of galaxies within the Coma Cluster exceeded what could be accounted for by the visible matter alone \cite{1933AcHPh...6..110Z} (review also \cite{2017arXiv171101693A}). He suggested that an additional, unseen mass—who coined the name dark matter— must be exerting gravitational influence to accelerate these orbital motions. Subsequent investigations developed by Ford and Rubin in the $1970$s \cite{APJ15970}, would show a consistent presence of unexplained high-speed orbits in every galaxy became apparent, solidifying the consensus around the existence of an exotic matter.

In the search for the exact nature of dark matter, to date, a large number of experiments have been carried out using a variety of theoretical models. Among the theoretical proposed candidates, we find weakly (WIMPs) \cite{2011PhRvD..83h3512K} and feebly (FIMP) \cite{2017IJMPA..3230023B,2010JHEP...03..080H} interacting massive particles, the neutralino \cite{2004PhRvD..69c5001H,2014JHEP...08..093H,rmfdm3}, Axions \cite{2009NJPh...11j5008D}, among others; however, no concrete conclusion has been reached about the nature of dark matter. Within this set of possibilities, in particular, the weakly interacting massive particles (WIMPs) are those more promising; review, for example, \cite{2018RvMP...90d5002B,pdg22}. 

Motivated by the detections of gravitational waves caused by the merger of a binary system recorded by the LIGO-Virgo collaboration (LVC) \cite{2016PhRvL.116v1101A,2017PhRvL.118v1101A,2017PhRvL.119n1101A,2017ApJ...851L..35A,2018PhRvL.121p1101A,2016ApJ...818L..22A,2016PhRvL.116x1103A,2016PhRvL.116f1102A,2017PhRvL.119p1101A,2017ApJ...848L..12A,pinku,2021PhRvD.103d3009K}, in recent years, numerous investigations have explored the integration of dark matter into hadronic matter \cite{rmfdm2,rmfdm13,Lopes_2018,karkevandi2,dmnosso1,rmfdm6,rmfdm8,rmfdm3,dmnosso2,abdul,rmfdm11,dmnosso3,laura-tolos,sagun,rmfdm10}, as well as the effects of dark matter in relation to quarks and hybrid stars \cite{fraga, lenziepjc2023} and on some properties of neutron stars, namely, how this type of exotic matter would influence some parameters macroscopic observations such as in the equilibrium configuration \cite{rmfdm2,rmfdm13,Lopes_2018,karkevandi2,dmnosso1,rmfdm6,rmfdm8}, radial stability \cite{pinku,2021PhRvD.103d3009K}, tidal deformability \cite{rmfdm13,rmfdm6,rmfdm3,dmnosso2,abdul,rmfdm11,dmnosso3,laura-tolos,sagun}, and $f_0$-mode frequency of oscillation into Cowling approximation \cite{rmfdm10}.


In the present work, we study the effect of dark matter on the stellar structure configuration, $f_0$-mode frequency of oscillation from complete general relativity, and tidal deformability. For the fluid inside the compact stars, we employ the relativistic mean field model with the NL3* parameterization \cite{snl3}. The contribution of dark matter is included through a kinetic term as done in \cite{dmnosso2,dmnosso3}, where we vary the Fermi momentum by taking different values of the neutralino mass, as detailed throughout the article. Once the equation of state (EOS) is defined, the mass, radius, tidal deformability, oscillation frequency, and damping time are analyzed for both hadronic stars and for hadronic stars in a binary system. These results are contrasted with observational data reported by LVC in the articles aforementioned. In addition, the $f_0$-mode derived from the complete general relativity is compared with the ones derived from the nonradial oscillation equations through the Cowling approximation.

The next sections are distributed as follows: in Sec.~\ref{EOS} the structure of the hadronic model with dark matter content is described and the main equations of state of the system are obtained. In Sec.~\ref{formalism} the equilibrium equation, nonradial oscillation equation, and tidal deformability equation are presented. In Sec.~\ref{results} we discuss the results and in Sec.~\ref{conclusion} we conclude the paper. Finally, throughout the text, we adopt the metric signature $(-,+,+,+)$ and we employ in geometrized units $c=1=G$.

\section{Hadronic model with dark matter content}
\label{EOS}

The impact of dark matter on the properties related to nonradial oscillations in neutron stars is performed here by considering the system, composed of hadronic matter admixed with DM, described by the following Lagrangian density
\begin{align}
\mathcal{L} = \mathcal{L}_{\had} + \mathcal{L}_{\dm},
\label{dlagtotal}
\end{align}
in which
\begin{align}
\mathcal{L}_{\had} &= \bar{\psi}(i\gamma^\mu\partial_\mu - M_{\mbox{\tiny nuc}})\psi + g_\sigma\sigma\bar{\psi}\psi 
- g_\omega\bar{\psi}\gamma^\mu\omega_\mu\psi
\nonumber \\ 
&- \frac{g_\rho}{2}\bar{\psi}\gamma^\mu\vec{b}_\mu\vec{\tau}\psi
+\frac{1}{2}(\partial^\mu \sigma \partial_\mu \sigma - m^2_\sigma\sigma^2)
- \frac{A}{3}\sigma^3 
\nonumber\\
& - \frac{B}{4}\sigma^4  -\frac{1}{4}F^{\mu\nu}F_{\mu\nu} 
+ \frac{1}{2}m^2_\omega\omega_\mu\omega^\mu 
-\frac{1}{4}\vec{B}^{\mu\nu}\vec{B}_{\mu\nu} 
\nonumber\\
&+ \frac{1}{2}m^2_\rho\vec{b}_\mu\vec{b}^\mu,
\label{dlaghad}
\end{align}
with nucleon and mesons fields denoted by $\psi$, $\sigma$, $\omega^\mu$, and $\vec{b}_\mu$, respectively (masses given by $M_{\mbox{\tiny nuc}}$, $m_\sigma$, $m_\omega$, and $m_\rho$). The tensors $F^{\mu\nu}$ and $\vec{B}^{\mu\nu}$ are defined as $F_{\mu\nu}=\partial_\mu\omega_\nu-\partial_\nu\omega_\mu$ and $\vec{B}_{\mu\nu}=\partial_\mu\vec{b}_\nu-\partial_\nu\vec{b}_\mu$. More details of this kind of relativistic mean-field (RMF) model applied to symmetric and asymmetric nuclear matter can be found, for instance, in Refs.~\cite{rev3,dutra2014}. The free parameters of the model are the couplings $g_\sigma$, $g_\omega$, $g_\rho$, $A$, $B$, and the particular set used here for these constants is the one given by the NL3* model~\cite{snl3}. Such parametrization was chosen due to its capability of reproducing ground state binding energies, charge radii, and giant monopole resonances of a set of spherical nuclei, namely, $^{16}\rm O$, $^{34}\rm Si$, $^{40}\rm Ca$, $^{48}\rm Ca$, $^{52}\rm Ca$, $^{54}\rm Ca$, $^{48}\rm Ni$, $^{56}\rm Ni$, $^{78}\rm Ni$, $^{90}\rm Zr$, $^{100}\rm Sn$, $^{132}\rm Sn$, and $^{208}\rm Pb$, as well as macroscopic properties of neutron stars. The complete study performed with more than 400 other parametrizations of the RMF model is found in Ref.~\cite{brett-jerome}. The dark sector of the model is given by~\cite{rmfdm13,rmfdm1,rmfdm2,rmfdm3,abdul,rmfdm6,rmfdm11,rmfdm10,rmfdm8,rmfdm12,dmnosso1,dmnosso2,dmnosso3}
\begin{align}
\mathcal{L}_{\dm} &= \bar{\chi}(i\gamma^\mu\partial_\mu - M_\chi)\chi
+ \xi h\bar{\chi}\chi +\frac{1}{2}(\partial^\mu h \partial_\mu h - m^2_h h^2)
\nonumber\\
&+ f\frac{M_{\mbox{\tiny nuc}}}{v}h\bar{\psi}\psi,
\label{dlagdm}
\end{align}
with the Dirac field $\chi$ representing the dark fermion of mass $M_\chi$. The scalar field $h$ denotes the Higgs boson with mass $m_h=125$~GeV, and the strength of the Higgs-nucleon interaction is controlled by $fM_{\mbox{\tiny nuc}}/v$, with $v=246$~GeV being the Higgs vacuum expectation value. The constant $\xi$ regulates the Higgs-dark particle coupling. 

The field equations of the system are calculated by using the mean-field approximation~\cite{rev3,dutra2014}, namely,
\begin{align}
m^2_\sigma\,\sigma &= g_\sigma \rho_s - A\sigma^2 - B\sigma^3, 
\\
m_\omega^2\,\omega_0 &= g_\omega \rho, 
\\
m_\rho^2\,b_{0(3)} &= \frac{g_\rho}{2}\rho_3, 
\\
[\gamma^\mu (&i\partial_\mu - g_\omega\omega_0 - g_\rho b_{0(3)}\tau_3/2) - M^*]\psi = 0,
\\
m^2_h\,h &= \xi \rho_s^{\dm} + f\frac{M_{\mbox{\tiny nuc}}}{v}\rho_s,
\label{hfield}
\\
(\gamma^\mu &i\partial_\mu - M_\chi^*)\chi = 0,
\end{align}
with $\tau_3=1$ for protons and $-1$ for neutrons, and effective dark particle and nucleon masses written as
\begin{eqnarray}
M^*_\chi = M_\chi - \xi h,
\end{eqnarray}
and
\begin{eqnarray}
M^* = M_{\mbox{\tiny nuc}} - g_\sigma\sigma - f\frac{M_{\mbox{\tiny nuc}}}{v}h,
\end{eqnarray}
respectively. The densities related to hadronic and dark sectors are 
\begin{align}
\rho_s &={\rho_s}_p+{\rho_s}_n,
\\
\rho &=\rho_p + \rho_n,
\\
\rho_3 &=\left<\bar{\psi}\gamma^0{\tau}_3\psi\right> = \rho_p - \rho_n =(2y_p-1)\rho, 
\\
\rho_{s_{p,n}} &= \left<\bar{\psi}_{p,n}\psi_{p,n}\right> =
\frac{M^*}{\pi^2}\int_0^{k_{F_{p,n}}} \hspace{-0.5cm}\frac{k^2dk}{(k^2+M^{*2})^{1/2}},
\\
\rho_s^{\dm} &= \left<\bar{\chi}\chi\right> = 
\frac{M^*_\chi}{\pi^2}\int_0^{k_F^{\dm}} \hspace{-0.5cm}\frac{k^2dk}{(k^2+M^{*2}_\chi)^{1/2}}.
\end{align}
The proton fraction is $y_p=\rho_p/\rho$, and the vector densities are related to the respective Fermi momenta through $\rho_{p,n}=2{k_F^3}_{p,n}/(3\pi^2)$, and $\rho_\chi=2{k_F^{\dm}}^3/(3\pi^2)$. In our study, we keep this last quantity as a free parameter, as well as the dark fermion mass, for which the values are taken from the range $50\mbox{ GeV}\leqslant M_\chi\leqslant 500\mbox{ GeV}$. Such interval for $M_\chi$, along with $\xi=0.01$, ensures that the spin-independent scattering cross-section is compatible with data provided by PandaX-II~\citep{panda}, LUX~\citep{lux}, and DarkSide~\citep{darkside} collaborations.

The energy density and pressure of the system is calculated from the energy-momentum tensor, $T^{\mu\nu}$, obtained through Eq.~\eqref{dlagtotal}. Such expressions are given, respectively, by
\begin{align} 
&\varepsilon(\sigma,\omega_0,b_{0(3)},\rho,y_p,h,\rho_\chi) = \frac{m_{\sigma}^{2} \sigma^{2}}{2} +\frac{A\sigma^{3}}{3} +\frac{B\sigma^{4}}{4} 
\nonumber\\
&-\frac{m_{\omega}^{2} \omega_{0}^{2}}{2} - \frac{m_{\rho}^{2} b_{0(3)}^{2}}{2} 
+ g_{\omega} \omega_{0} \rho + \frac{g_{\rho}}{2} b_{0(3)}\rho_3 + \frac{m_h^2h^2}{2} 
\nonumber\\
&+ \varepsilon_{\kin}^{p} + \varepsilon_{\kin}^{n} + \varepsilon_{\kin}^{\dm},
\label{eden}
\end{align}
and
\begin{align}
&P(\sigma,\omega_0,b_{0(3)},\rho,y_p,h,\rho_\chi) = -\frac{m_{\sigma}^{2} \sigma^{2}}{2} - \frac{A\sigma^{3}}{3} - \frac{B\sigma^{4}}{4} 
\nonumber\\
&+ \frac{m_{\omega}^{2} \omega_{0}^{2}}{2} + \frac{m_{\rho}^{2} b_{0(3)}^{2}}{2} - \frac{m_h^2h^2}{2} + P_{\kin}^p + P_{\kin}^n + P_{\kin}^{\dm},
\label{press}
\end{align}
with the kinetic terms written as
\begin{eqnarray}
\varepsilon_{\kin}^{\dm} &=& \frac{1}{\pi^2}\int_0^{k_F^{\dm}}
\hspace{-0.3cm}k^2(k^2+M^{*2}_\chi)^{1/2}dk,
\label{ekindm}
\\
P_{\kin}^{\dm} &=& 
\frac{1}{3\pi^2}\int_0^{{k_F^{\dm}}}\hspace{-0.5cm}\frac{k^4dk}{(k^2+M^{*2}_\chi)^{1/2}},
\label{pkindm}
\\
\epsilon_{\kin}^{p,n} &=& \frac{1}{\pi^2} \int_0^{{k_F}_{p,n}}k^2({k^{2}+M^{*2}})^{1/2}dk,
\\
P_{\kin}^{p,n} &=&  
\frac{1}{3\pi^2} \int_0^{k_{F\,{n,p}}}\hspace{-0.5cm}  
\frac{k^4dk}{\left({k^{2}+M^{*2}}\right)^{1/2}}.
\end{eqnarray}
In these expressions, $\sigma$, $\omega_0$, $b_{0(3)}$, and $h$ are the mean-field values of the respective mesonic fields of the model. 

We also include electrons and muons in the system in order to correctly describe the core of neutron stars. Therefore, energy density and pressure become
\begin{align}
\epsilon_{\rm core} &= \varepsilon(\sigma,\omega_0,b_{0(3)},\rho,y_p,h,\rho_\chi) + \frac{\mu_e^4}{4\pi^2}
\nonumber\\
&+ \frac{1}{\pi^2}\int_0^{\sqrt{\mu_\mu^2-m^2_\mu}}\hspace{-0.6cm}dk\,k^2(k^2+m_\mu^2)^{1/2},
\end{align}
and
\begin{align}
p_{\rm core} &= P(\sigma,\omega_0,b_{0(3)},\rho,y_p,h,\rho_\chi) + \frac{\mu_e^4}{12\pi^2}
\nonumber\\
&+\frac{1}{3\pi^2}\int_0^{\sqrt{\mu_\mu^2-m^2_\mu}}\hspace{-0.5cm}\frac{dk\,k^4}{(k^2+m_\mu^2)^{1/2}}.
\end{align}
The last terms of the above equations represent the thermodynamical quantities of massless electrons, and muons with mass $m_\mu=105.7$~MeV. The chemical potentials of these leptons, denoted by $\mu_e$ and $\mu_\mu$, are related to their respective densities through $\rho_e=\mu_e^3/(3\pi^2)$, and \mbox{$\rho_\mu=[(\mu_\mu^2 - m_\mu^2)^{3/2}]/(3\pi^2)$}. For a beta-equilibrated system submitted to charge neutrality, the following conditions apply, namely, $\rho_p - \rho_e  = \rho_\mu$, and $\mu_n - \mu_p = \mu_e=\mu_\mu$. Finally, we consider two different regions for the description of the neutron star crust: the outer (OC) and the inner crust (IC). For the former we use the equations proposed by Baym, Pethick and Sutherland (BPS)~\cite{bps} in a density region of 
$6.3\times10^{-12}\,\mbox{fm}^{-3}\leqslant\rho\leqslant2.5\times10^{-4}\,\mbox{fm}^ {-3}$. For the inner crust we use a polytropic relation between energy density and pressure, namely, $p_{\rm IC}(\epsilon_{\rm IC})=A+B\epsilon_{\rm IC}^{4/3}$. We match this form to the BPS, and to the core equations. The latter is connected at the the core-crust transition pressure and energy density, calculated through the thermodynamical method~\cite{tm1,tm2,tm4}. Total energy density and total pressure of the stellar matter is then given by 
\begin{eqnarray}
\epsilon = \epsilon_{\rm core} + \epsilon_{\rm OC} + \epsilon_{\rm IC}  
\end{eqnarray}
and 
\begin{eqnarray}
p = p_{\rm core} + p_{\rm OC} + p_{\rm IC},
\end{eqnarray}
respectively.


\section{General relativistic formalism}\label{formalism}

\subsection{The stellar equilibrium equations}

The static equilibrium configurations of a compact star composed of a perfect fluid {are obtained through} the Tolman-Oppenheimer-Volkoff equations
\begin{eqnarray}
\label{tov1}
&&\frac{dp}{dr} = - \frac{\epsilon m}{r^2}\bigg(1 + \frac{p}{\epsilon}\bigg)
	\bigg(\frac{1 + \frac{4\pi p r^3}{m}}{1 - \frac{2m}{r}}\bigg),  \\
\label{tov2}
&&\frac{d\nu}{dr} = - \frac{2}{\epsilon} \frac{dp}{dr} \bigg(1 + \frac{p}{\epsilon}\bigg)^{-1}, \\
\label{tov3}
&&\frac{dm}{dr}= 4 \pi r^2 \epsilon.
\end{eqnarray}
{The variables $m(r)$ and $\nu(r)$ are respectively the gravitational mass inside the radius $r$ and a metric potential.} The pressure $p$ and the mass-energy density $\epsilon$ are {connected} by the equations of state. To integrate the system of Eqs. \eqref{tov1}-\eqref{tov3} from the center ($r=0$) to the star's surface ($r=R$), some conditions are required. At the center of the star ($r=0$), we have
\begin{eqnarray}
\epsilon(0)=\epsilon_c,\quad p(0)=p_c,\quad {\rm and}\quad m(0)=0.   
\end{eqnarray}
The surface of the star is found when $p(R)=0$. At this point
\begin{eqnarray}
    \nu(R)= \ln\left( 1- \frac{2M}{R}\right),
\end{eqnarray}
with $M$ representing the total stellar mass.

\subsection{The nonradial oscillation equations}

{To investigate the non-radial oscillation, both spacetime and fluid variables are perturbed. The perturbations are replaced in the Einstein field equation, in the energy-momentum tensor, and in the baryon number conservation maintaining only the first-order variables.} 

{Following  Ref. \cite{1983ApJS...53...73L, 1985ApJ...292...12D}, we adopt the perturbed line element of the form}
%
\begin{eqnarray}
ds^2 & = & -e^{\nu}(1+r^{\ell}H_0Y^{\ell}_{m}e^{i\omega t})dt^2  - 2i\omega r^{\ell+1}H_1Y^{\ell}_me^{i\omega t}dtdr   \nonumber \\
&& + e^{\lambda}(1 - r^{\ell}H_0Y^{\ell}_{m}e^{i \omega t})dr^2 \nonumber \\
&& + r^2(1 - r^{\ell}KY^{\ell}_{m}e^{i \omega t})(d\theta^2 + \sin^2\theta d\phi^2),
\end{eqnarray}
{with $H_0=H_0(r)$, $H_1=H_1(r)$, and $K=K(r)$ being functions of the radial coordinate $r$ only, $\omega$ depicting the eigenfrequency of oscillation, and $Y^{\ell}_{m}=Y^{\ell}_{m}(\theta,\phi)$ representing the even-parity spherical harmonic functions.}
%
%
{Since the small perturbations are set by the Lagrangian fluid displacement $\xi^{\beta}$, we consider:}
\begin{eqnarray}
&&\xi^{r} = r^{\ell-1}e^{-\lambda/2}WY^{\ell}_{m}e^{i\omega t}, \\
&&\xi^{\theta} = -r^{\ell - 2}V\partial_{\theta}Y^{\ell}_{m}e^{i\omega t}, \\
&&\xi^{\phi} = -r^{\ell}(r \sin \theta)^{-2}V\partial_{\phi}Y^{\ell}_{m}e^{i\omega t} ,
\end{eqnarray}
{where $\xi^{0}=0$.} 

With this, non-radial oscillations are described by the following set of first-order linear differential equations \cite{1985ApJ...292...12D}:
\begin{eqnarray}
H_1' &=&  -r^{-1} \left[ \ell+1+ \frac{2Me^{\lambda}}{r} +4\pi   r^2 e^{\lambda}(p-\epsilon)\right] H_{1}   \nonumber \\ 
&& +  e^{\lambda}r^{-1}  \left[ H_0 + K - 16\pi(\epsilon+p)V \right] ,      \label{osc_eq_1}  \\
 K' &=&    r^{-1} H_0 + \frac{\ell(\ell+1)}{2r} H_1   - \left[ \frac{(\ell+1)}{r}  - \frac{\nu'}{2} \right] K  \nonumber \\
&&  - 8\pi(\epsilon+p) e^{\lambda/2}r^{-1} W \:,  \label{osc_eq_2} \\
 W' &=&  - (\ell+1)r^{-1} W   + r e^{\lambda/2} \left[ e^{-\nu/2} \gamma^{-1}p^{-1} X\right.    \nonumber \\
&& \left.- \ell(\ell+1)r^{-2} V + \frac{1}{2}H_0 + K \right] \:,  \label{osc_eq_3}
\\
X' &=&  - \ell r^{-1} X + \frac{(\epsilon+p)e^{\nu/2}}{2}  \Bigg[ \left( \frac{1}{r}+\frac{\nu'}{2} \right)H_{0}  \nonumber  \\
&& + \left(r\omega^2e^{-\nu} + \frac{\ell(\ell+1)}{2r} \right) H_1    + \left(\frac{3}{2}\nu' - \frac{1}{r}\right) K   \nonumber  \\
&&    - \ell(\ell+1)r^{-2}\nu' V  -  2 r^{-1}   \Biggl( 4\pi(\epsilon+p)e^{\lambda/2} \nonumber  \\
&& + \omega^2e^{\lambda/2-\nu}  - \frac{r^2}{2}  (e^{-\lambda/2}r^{-2}\nu')' \Biggr) W \Bigg]  \:,
\label{osc_eq_4}
\end{eqnarray}
where the prime {stands} a derivative with respect to $r$ and  $\gamma$ {depicts} the adiabatic index. The function $X$ is given by
\begin{eqnarray}
X =  \omega^2(\epsilon+p)e^{-\nu/2}V - \frac{p'}{r}e^{(\nu-\lambda)/2}W   \nonumber \\
  + \frac{1}{2}(\epsilon+p)e^{\nu/2}H_0 ,
\end{eqnarray}
{with} $H_{0}$ {fulfilling} the algebraic relation 
\begin{eqnarray}
a_1  H_{0}= a_2  X -  a_3 H_{1}  + a_4 K,
\end{eqnarray}
{where}
\begin{eqnarray}
a_1 &=&  3M + \frac{1}{2}(l+2)(l-1)r + 4\pi r^{3}p  ,  \\
a_2 &=&  8\pi r^{3}e^{-\nu /2}   , \\
a_3 &=&  \frac{1}{2}l(l+1)(M+4\pi r^{3}p)-\omega^2 r^{3}e^{-(\lambda+\nu)}  ,  \\
a_4 &=&  \frac{1}{2}(l+2)(l-1)r - \omega^{2} r^{3}e^{-\nu}  \nonumber \\
    &&    -r^{-1}e^{\lambda}(M+4\pi r^{3}p)(3M - r + 4\pi r^{3}p)   .
\end{eqnarray}
Outside the {stellar structure configuration}, the perturbation functions that describe the motion of the fluid {$W$ and $V$} vanish, and the system of differential equations reduces to the Zerilli equation:
\begin{equation}
\frac{d^{2}Z}{dr^{*2}}=[V_{Z}(r^{*})-\omega^{2}]Z,
\end{equation}
where $Z(r^{*})$ and $dZ(r^{*})/dr^{*}$ are related to the {functions} $H_{0}(r)$ and $K(r)$ {through the} transformations given in Refs. \cite{1983ApJS...53...73L,1985ApJ...292...12D}. {$Z(r^{*})$ depends on} the ``tortoise'' coordinate {which is given by}
\begin{equation}
    r^{*} = r + 2 M \ln\left(\frac{r}{2M} -1\right),
\end{equation}
and the effective potential $V_{Z}(r^{*})$ {yields}
\begin{eqnarray}
V_{Z}(r^{*}) = \frac{(1-2M/r)}{r^{3}(nr + 3M)^{2}}[2n^{2}(n+1)r^{3} + 6n^{2}Mr^{2}  \nonumber \\
 + 18nM^{2}r + 18M^{3}],
\end{eqnarray}
with $n= (l-1) (l+2) / 2$.

{For given values of $l$ and $\omega$, the set of Eqs. (\ref{osc_eq_1})$-$(\ref{osc_eq_4}) has four linearly independent solutions.} The physical solution {obtained must verify} boundary conditions: (a) The perturbation functions {have to be regular throughout the entire star; thus implying that in $r = 0$ must also be finite, since at this point non-radial oscillation equations are singular.} To {insert} such condition, {the solution near $r=0$ must be expanded using a power series (for more detail about the process, see Ref. \cite{1983ApJS...53...73L}.} (b) {Since the} Lagrangian perturbation {of} pressure {vanishes at the star's surface} $r = R$. {This boundary condition is equivalent to $X(R)=0$.} {For a set of values of} $l$ and $\omega$, there is a solution that {fulfills} the above boundary conditions inside the star.

In general, {in the exterior of the star,} the perturbed {red}{line element depicts a mixture} of outgoing and ingoing gravitational waves. {Due to we focus on the purely outgoing gravitational radiation at $r=\infty$, the Zerilli equation is used. The eigenfrequency of oscillation that satisfies this requirement represents the quasinormal modes of the stellar model.} {process} to solve the above equations is {detailed} in \cite{1983ApJS...53...73L, 1985ApJ...292...12D}.

{Different oscillation} modes can be {investigated using the} {aforementioned} equations. The purely gravitational modes ($w$-modes) do not induce fluid motion and are highly damped. The fluid {pulsation} modes ($f_0$, $g$, and $p$-modes) are usually {grouped} according to the {origin} of the restoring force which prevails in bringing the perturbed element of fluid back to the equilibrium position; e.g., {the} buoyancy in the case of $g$-modes or a gradient of pressure for $p$-modes. 
{It is widely known that the frequencies of the} $g$-modes are lower than those of $p$-modes, and {these} two sets are separated by the frequency of the $f_0$-mode. 
The $f_0$-mode frequency is proportional to the square root of the mean density of the star and tends to be independent of the details of the stellar structure. 
In this work, we {analyze} only on the $f_0$-mode because it is expected to be the most excited in astrophysical events and therefore the main contribution to the GW emission of the star should be expected at this frequency.

\begin{figure}[tb]
 \centering
 \includegraphics[trim=0cm 6cm 0cm 6cm, clip, scale=0.41]{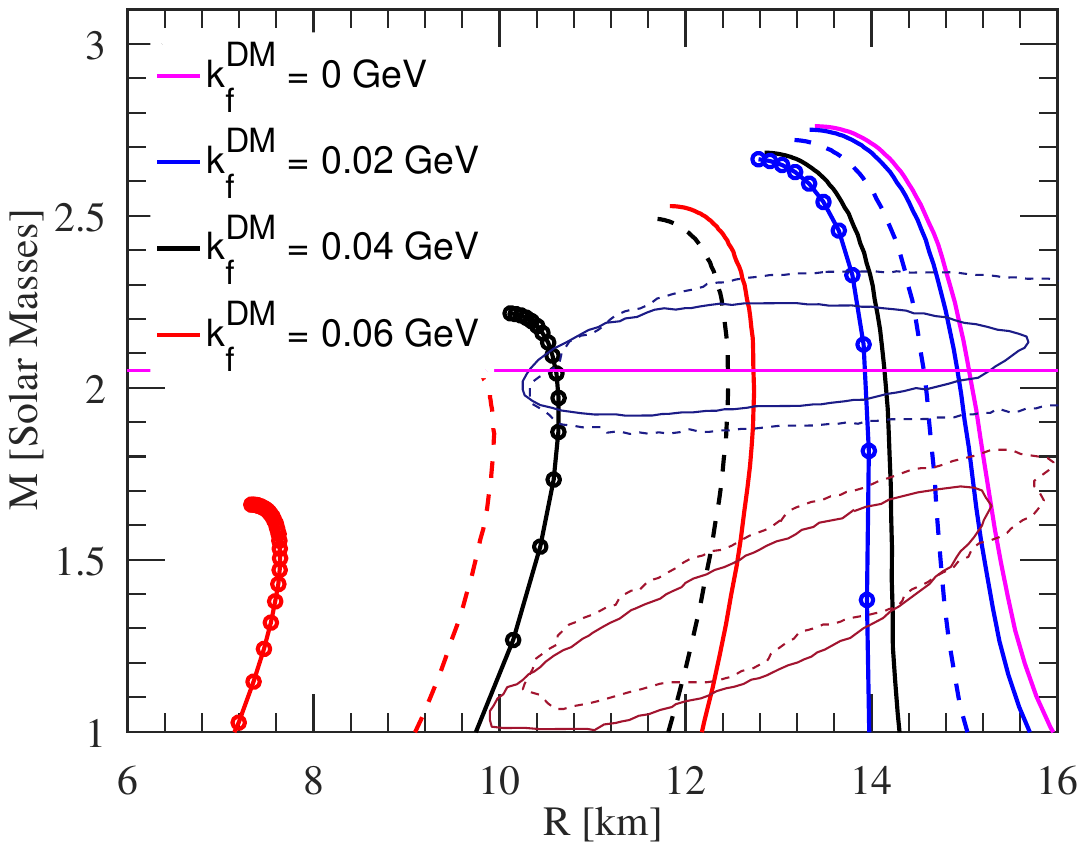} 
\caption{Mass{, normalized in Sun's masses,} versus radius for {some values of $M_{\chi}$ and $k_F^{\dm}$}. {The curve in magenta represents the case without dark matter. The curves of the same color represent the results for a fixed value of $k_F^{\dm}$ and the curves of the same type depict results for a fixed value of $M_{\chi}$.} {The} solid curves correspond to $M_{\chi} = 50$~MeV, dashed curves correspond to $M_{\chi} = 200$~MeV, and solid curves with circles correspond to the case $M_{\chi} = 500$~MeV. Red and blue regions represent the masses and radii intervals (95\%) from  PSR J$0030+0451$ and PSR J$0740+6620$ measured by NICER~\cite{Riley:2019yda, Riley:2021pdl, Miller:2021qha, Miller:2019cac}. The magenta horizontal line includes all observed neutron stars over 2 solar masses, which include the pulsars PSR J$1614-2230$, PSR J$0348+0432$, and PSR J$0740+6620$~\cite{Demorest:2010bx,Antoniadis:2013pzd,Cromartie:2019kug}}.
\label{fig1}
\end{figure}

\subsection{The tidal deformability equations}

{The analysis of tidal deformability is commonly investigated in binary compact object systems. The gravitational effects caused by one star can bring out the deformation of its partner. This deformation was investigated} by Damour and Nagar, Binnington and Poisson \cite{PhysRevD.80.084035,PhysRevD.80.084018}. They {found} that the tidal deformation of a neutron star is {determined} by the gravito-electric $K^{el}_{2}$ and gravito-magnetic $K^{mag}_{2}$ Love numbers, {which are respectively connected with} the mass quadrupole and {with} the current quadrupole induced by the companion star. {Additional research carried out by} Flanagan and Hindeler {found} that a single detection should be sufficient to impose upper limits on $K^{el}_{2}$ at 90\% confidence level \cite{PhysRevD.77.021502}. Since then, {intensive} research has been {carried out} on the {calculation of the} Love numbers of neutron stars
\cite{PhysRevC.87.015806,PhysRevC.98.035804,PhysRevC.98.065804,PhysRevC.95.015801,2010PhRvD..81l3016H}.  

In a binary {stellar system,} the induced quadrupole moment $Q_{ij}$ in one neutron star {because of} the external tidal field ${\cal E}_{ij}$ {induced} by a {partner} compact object can be {expressed} as \cite{2010PhRvD..81l3016H}
\begin {equation}
Q_{ij} = -\lambda {\cal E}_{ij},
\end{equation}
where $\lambda$ is the tidal deformability parameter, which can be {written} in terms of {the} quadrupole {($l = 2$)} tidal Love number $k_2$ as
\begin{equation}
\label{tidal}
\lambda= \frac{2}{3} {k_2}R^{5}.
\end{equation}
To {determine} $k_{2}$ we {need} to solve the {first-order} differential equation  
\begin{equation}
r \frac{dy}{dr} + y^2 + y F(r) + r^2Q(r)=0,
\label{ydef}
\end{equation}
{with the functions $F(r)$ and $Q(r)$} given by
\begin{eqnarray}
&&F(r)= [1 - 4\pi r^2(\varepsilon - p)]/E\\
&&Q(r)=4\pi \left[5\varepsilon + 9p + (\varepsilon + p)\left(\frac{\partial 
p}{\partial\varepsilon}\right)-\frac{6}{4\pi r^2}\right]/E 
\nonumber\\ 
&&- 4\left[ \frac{m+4\pi r^3 p}{r^2 E} \right]^2,
\end{eqnarray}
with $E = 1-2m/r$. Therefore the Love number $k_2$ {is computed to be}
\begin{align}
&k_2 =\frac{8C^5}{5}(1-2C)^2[2+C(y_R-1)-y_R]\times
\nonumber\\
&\times\Big\{2C [6-3y_R+3C(5y_R-8)]
\nonumber\\
&+4C^3[13-11y_R+C(3y_R-2) + 2C^2(1+y_R)]
\nonumber\\
&+3(1-2C^2)[2-y_R+2C(y_R-1)]{\rm ln}(1-2C)\Big\}^{-1},
\label{k2}
\end{align}
where {the function} $y_R = y(r = R)$ and $C= M/R$ {is the compactness parameter}. Equation (\ref{ydef}) has to be solved coupled {with} the TOV equations.

The dimensionless tidal deformability $\Lambda$ (i.e., the dimensionless version of $\lambda$) is connected with the {Love number} through the relation 
\begin{equation}
\Lambda= \frac{2k_2}{3C^5}.
\label{dtidal}
\end{equation}

\begin{figure}[tb]
\centering
\includegraphics[trim=0cm 6cm 0cm 6cm, clip, scale=0.41]{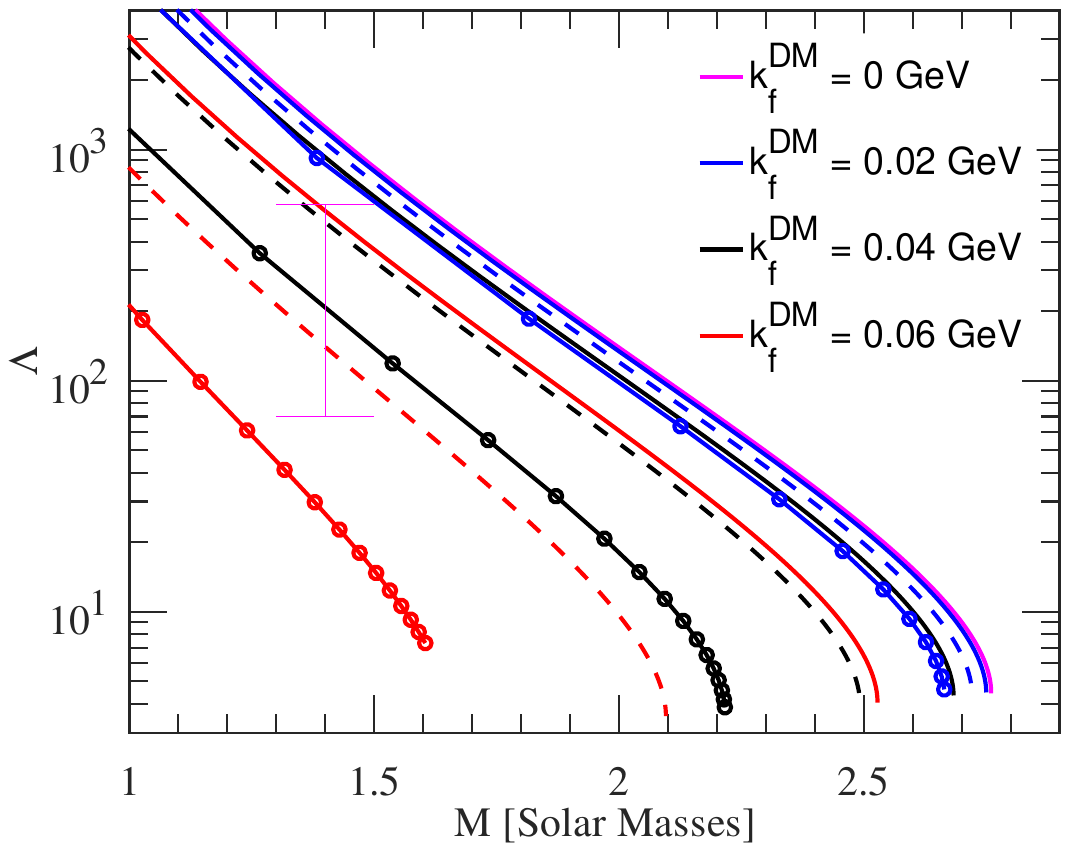} 
\caption{{Tidal deformability} versus mass {different values of $M_{\chi}$ and $k_F^{\dm}$. The curve in magenta stands for the case without dark matter. The curves with the same type of color represent results obtained with the same value of $k_F^{\dm}$ and the curves of the same type depict the results found with the same value $M_{\chi}$.} {The} solid curves correspond to $M_{\chi} = 50$~MeV, dashed curves correspond to $M_{\chi} = 200$~MeV, and solid curves with circles correspond to the case $M_{\chi} = 500$~MeV. {The vertical solid line depicts $\Lambda_{1.4}=190_{-120}^{+390}$ published in Ref. \cite{2017ApJ...848L..12A}.}}
\label{fig2}
\end{figure}

\section{Results}\label{results}

\begin{table}[ht]
\footnotesize
\setlength{\tabcolsep}{3pt} 
\renewcommand{\arraystretch}{0.8} 
\begin{tabular}{|c|c|c|c|c|c|} 
\toprule
$k_F^{\dm}$ &  $M_{\chi}$ & $M_{\rm max}$ & $R$ & $f_{0}$ & $\tau$ \\
(GeV)      &  (MeV)        & ($M_\odot$) & (km)    &  (kHz)       &   (s)  \\ 
\toprule
0.00 &  -  &  2.760 &  13.39 & 1.930 & 1.940 \\ \hline
\ \  &  50  & 2.750 & 13.33 & 1.920  & 1.920 \\
0.02 &  200  & 2.720 & 3.110 & 1.980  & 1.940 \\
\ \  &  500  & 2.670 & 12.72 & 2.000 & 1.870 \\\hline

\ \  &  50  & 2.680 & 12.85 & 1.990  & 1.890 \\
0.04 &  200  & 2.690 & 11.61 & 2.150 & 1.780 \\
\ \  &  500  & 2.220 & 10.08 & 3.280 & 1.230 \\ \hline

\ \  &  50  & 2.520 & 11.83 & 2.120  & 1.780 \\
0.06 &  200  & 2.100 & 9.450 & 2.580  & 1.520 \\
\ \  &  500  & 1.660 & 7.330 & 3.280 & 1.230 \\

\toprule 
\end{tabular}
\caption{Maximum mass values $M_{\text{\rm max}}$, and {their} corresponding radius $R$, fundamental mode frequency $f_{0}$, and damping time of the fundamental mode $\tau$, calculated for different values of fermi moment of dark matter $k_F^{\dm}$ and dark matter mass $M_{\chi}$.}
\label{table1}
\end{table}


\subsection{General remarks}

The full general relativistic equations for the oscillations of neutron stars (when the fluid and spacetime variables are perturbed), where the energy-momentum tensor is coupled with gravitational radiation (for $l=2$, i.e. quadrupole oscillations), are treated in the following subsections. In addition, there is a well-known framework called the Cowling approximation, where it is considered that the fluid perturbations are weakly coupled with the spacetime vibrations, so in this case we can consider only a set of differential equations for the fluid variables.

In the case of results obtained by using the Cowling approximation, we employ a code that implements the shooting method. In this procedure, there exist boundary conditions to be considered in the center and on the surface of the star. In the first stage, in the center of the star, we consider the initial values of the functions $W$ and $V$ (which satisfy regularity conditions), and we also introduce a test value for the $f_0$-mode frequency, which is typically around $2.5$ kHz. Of course, this proof value is suitable for a standard neutron star. After, we proceed with the numerical integration which starts in the center and ends on the surface of the star; this numerical interaction is developed through the Runge Kutta method. After finishing the integration process, we arrive at the surface of the star. At this point, we use the Newton-Raphson method to see if the estimated frequency value (the value of the eigenfunction) satisfies the boundary condition. If we do not find a good precision, the test value for the frequency is corrected and all the process is initialized again, until reach the desired precision for the oscillation mode, which will be the desired frequency. Finally, we have to mention that the numerical estimation considering the Cowling approximation has two main objectives; the first one is the direct calculation of the frequency of the $f_0$-mode and the second one is to use this value as the best proof value for the numerical integration of the full version of the general relativistic oscillation equations. This numerical code allows us to obtain the $f_0$-mode reported in literature, e.g., it reproduces the results of Ref. \cite{2023PhRvD.107l4016A}.

\begin{figure*}[tb]
 \centering
\includegraphics[trim=0cm 6cm 0cm 6cm, clip, scale=0.41]{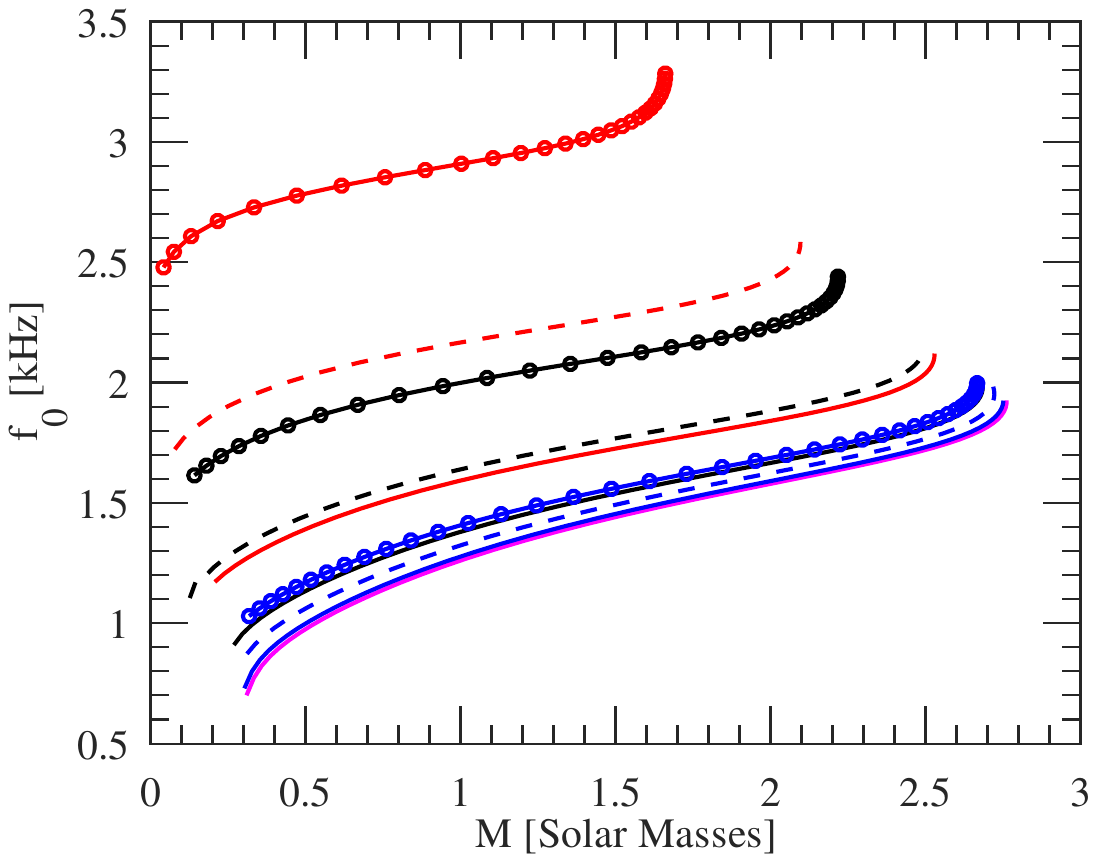}
\includegraphics[trim=0cm 6cm 0cm 6cm, clip, scale=0.41]{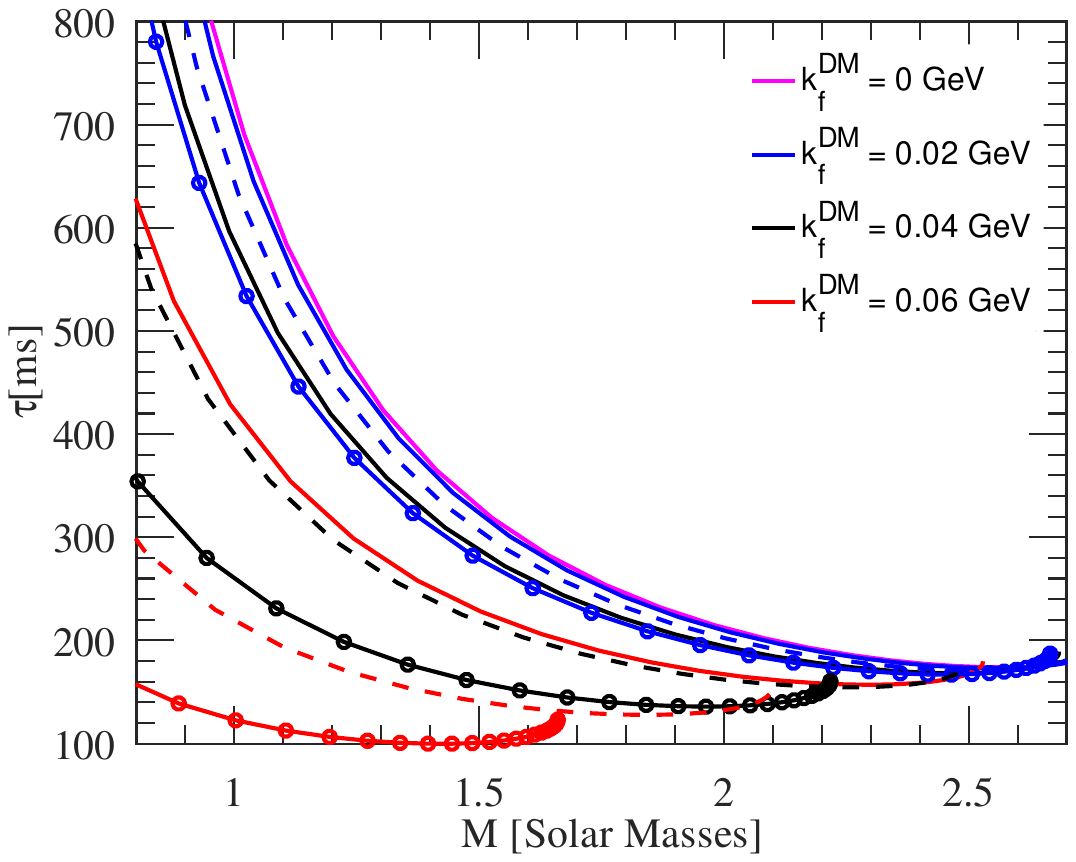}
\includegraphics[trim=0cm 6cm 0cm 6cm, clip, scale=0.41]{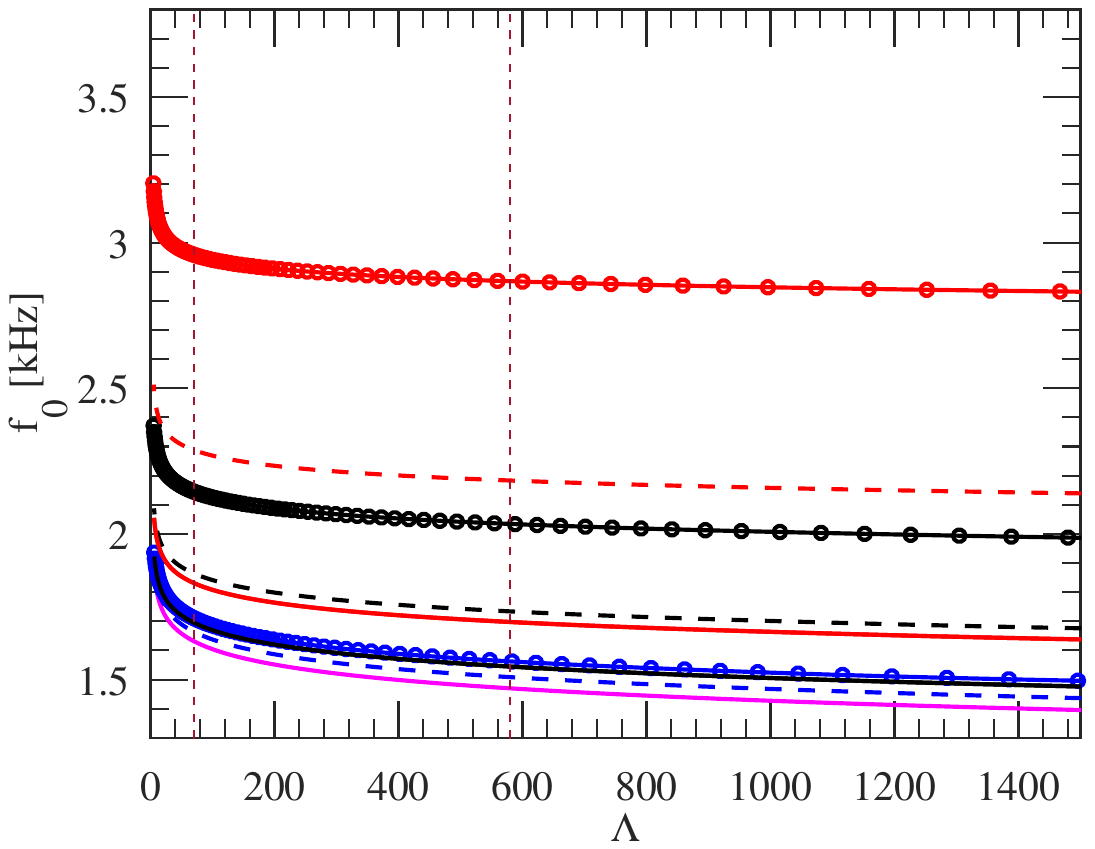} 
\includegraphics[trim=0cm 6cm 0cm 6cm, clip, scale=0.41]{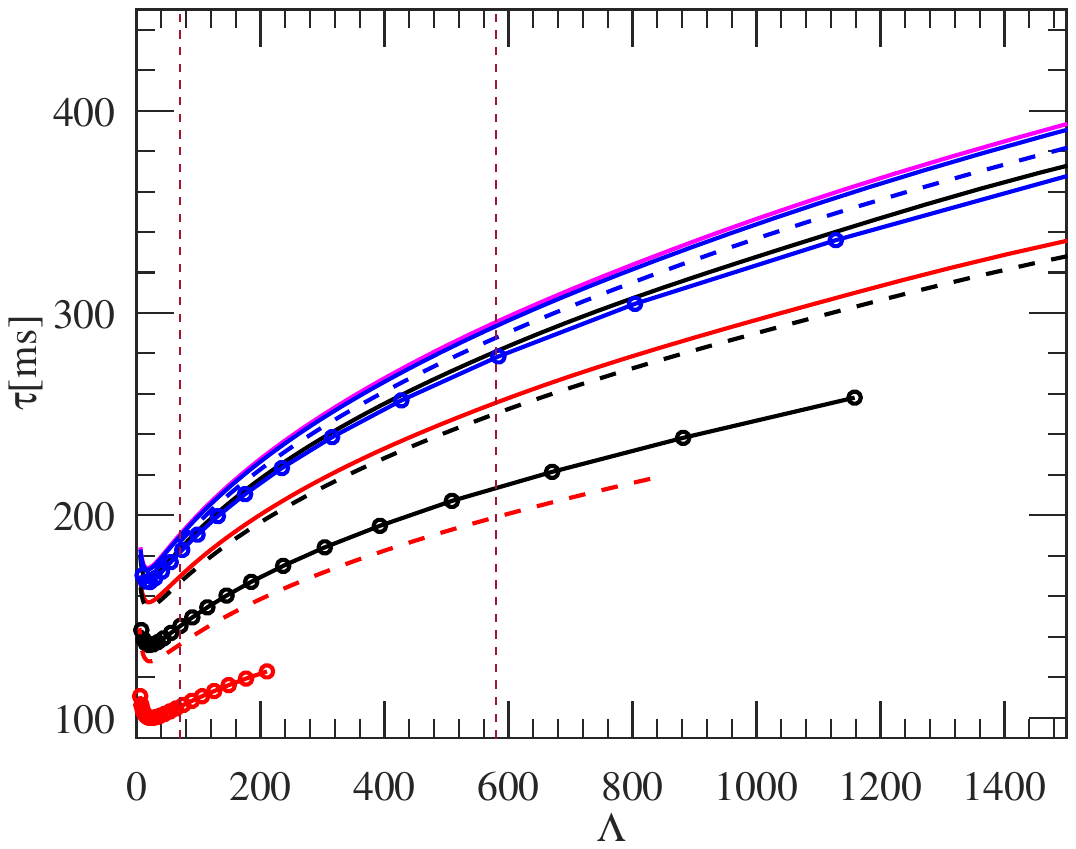} 
\caption{Fundamental mode frequency $f_0$ and damping time of fundamental mode $\tau$ against the total mass $M/M_{\odot}$ are shown on the top panel and versus the tidal deformability $\Lambda$ on the bottom panel. In all cases, different values of $M_{\chi}$ and $k_F^{\dm}$ are employed. The curves in magenta depict the case without dark matter. The curves of the same color represent the results for a fixed value of $k_F^{\dm}$ and the curves of the same type stand results for a fixed value of $M_{\chi}$. The solid curves correspond to $M_{\chi} = 50$~MeV, dashed curves correspond to $M_{\chi} = 200$~MeV, and solid curves with circles correspond to the case $M_{\chi} = 500$~MeV. In the bottom panels, the dashed vertical lines represent the tidal deformability obtained from the event GW$170817$, i.e., $\Lambda_{1.4}=190^{+390}_{-120}$, reported Ref. \cite{2017ApJ...848L..12A}.
} 
\label{fig3}
\end{figure*}

In a more rigorous treatment of neutron star seismology, we have to consider the coupling between the fluid movement and the metric perturbations. In this case, the numerical procedure is a very elaborate and long process of shooting method. The first step is to use as our best test value for the gravitational wave frequency the one obtained in the framework of the Cowling approximation, i.e., $\omega_{Cow}$. With such proof value, we begin the integration inside the star, from the center towards the star's surface. At first, it is selected three linearly independent solutions compatible with the regularity conditions in the center of the star, and then we begin with the numerical integration from $r = 0$ to the point $R/2$ inside the star. After that, a second integration is realized, where we select two linearly independent solutions compatible with the boundary condition on the star's surface, this process is made from the surface $R$ to the point $R/2$. To complete the procedure inside the star, the five solutions are combined to obtain compatibility with the boundary conditions at the center and on the surface of the star. Thereafter, we continue with the integration outside the star. The boundary values for the Zerilli function and its derivative at the surface of the star can always be obtained from the values of $H_0(R)$ and $K(R)$ obtained from the integration inside the compact star. Then, $Z(r^{*})$ can be determined outside the star by integrating the Zerilli equation. At an asymptotically large radius (which represents the infinity), the two linearly independent solutions of the Zerilli equation may be expressed as power series
\begin{equation}
Z_{-}(r^{*}) = e^{-i \omega r^{*}} \sum_{j=0}^{\infty} \beta_{j}r^{-j}    
\end{equation}
and
\begin{equation}
Z_{+}(r^{*}) = e^{i \omega r^{*}} \sum_{j=0}^{\infty} \bar{\beta}_{j}    
\end{equation}
where $Z_{-}$ depicts the radiated waves and $Z_{+}$ the ingoing waves. The general asymptotic solution is given by the linear combination
\begin{equation}
    Z(r^{*}) = A(\omega)Z_{-}(r^{*}) + B(\omega)Z_{+}(r^{*}),
\end{equation}
with $A(\omega)$ and $B(\omega)$ being complex numbers. The integration of the Zerilli differential equation from the surface of the star to the infinity ($r_{\infty} \approx 55 \omega^{-1}$) allows finding $Z$ and $dZ/dr$. Then, by using $Z(r^{*})$ at $r^{*} = r^{*}_{\infty}$, we can obtain $B(\omega)=0$ against to $Z(r^{*}_{\infty})$ and $dZ/dr(r^{*}_{\infty})$ for each value of $\omega$. The boundary condition at infinity requires that we only have outgoing gravitational radiation, i.e., to find the frequencies of quasi-normal modes we must obtain the roots of $B(\omega) = 0 $. This is achieved by determining $B(\omega)$ for three different close trial values of the eigenfrequency $\omega$. Then, we fit a quadratic polynomial $B(\omega) = \gamma_{0} + \gamma_{1}\omega + \gamma_{2}\omega^{2}$  to the computed values of $\omega$ to obtain an approximate root of $B(\omega) = 0 $. Finally, we iterate this procedure using the real part of the approximate root as an input for the next integration of the oscillation equations. This interaction is repeated until the real part value of the quasi-normal mode changes from one step to the next by less than one part in 10$^{8}$. The imaginary part of $\omega$ is connected to the gravitational damping time of the oscillation $\tau$. The numerical code implemented in this case allows us to reproduce the results shown in the reference \cite{2019JCAP...06..051V}.

In this work, we {analyze the effects} of dark matter inside neutron stars. {Since dark matter would interact very weakly with normal matter, its possible effects just could be felt only in a gravitational field.}

For our purposes, the dark matter mass $M_{\chi}$ and the Fermi momentum of dark matter particles $k_F^{\dm}$ values were selected to satisfy the restrictions obtained of both stellar mass and the tidal deformability of neutron stars observed; namely, we use $M_{\chi}=50, 200$, and $500$~MeV and $k_F^{\dm}=0, 0.02, 0.04$, and $0.06$~GeV. As a consequence, by using the NL3* EoS, we reproduce the mass, radius, $f_0$-mode frequency, damping time of the fundamental mode, and tidal deformability.

\subsection{Tidal deformability, frequency of oscillations, and damping time of fundamental mode for a hadronic star}

{From} Fig. \ref{fig1}, {the mass, normalized in solar masses, as a function of the total radius is plotted for some values of $M_{\chi}$ and $k_F^{\dm}$. The observation data belong to the NICER restrictions obtained from the pulsars PSR J$0030+0451$ \cite{Riley:2019yda,Miller:2019cac} and PSR J$0740+6620$ \cite{Riley:2021pdl, Miller:2021qha}. There are also shown the bands of the pulsars PSR J$0740+6620$ \cite{Cromartie:2019kug}, PSR J$0348+0432$ \cite{Antoniadis:2013pzd}, and PSR J$1614+2230$ \cite{Demorest:2010bx}. In the figure,} we observe that there {is} a strong dependence of the stellar mass on the parameters {$M_{\chi}$ and} $k_F^{\dm}$. For {a fixed $M_{\chi}$}, {an increase} of $k_F^{\dm}$ bring as a consequence {a decrease of} the maximum mass. It means that the {increase} in the fermi momentum of dark matter {produces a softness} in the EOS, therefore {there is a lower pressure, thus obtaining an equilibrium configuration with lower masses}. {In the same figure}, we also observe that an increase in $M_{\chi}$ {induces} a decrease in the stellar mass; {i.e.}, we have a softer EOS {when that parameter is increased}. {From these results, we observe that the change of $M_{\chi}$ and $k_F^{\dm}$ allows us to find some results more precise and closer to empirical evidence of neutron stars PSR J$0030+0451$, PSR J$0740+6620$, PSR J$0740+6620$, PSR J$0348+0432$, and PSR J$1614+2230$.}

{The deformation of a compact star within a binary system is} measured by the tidal deformability $\Lambda$. {In this way,} Fig. \ref{fig2} {shows} the deformability parameter as a function of the mass. {These results are compared with the case $\Lambda_{1.4}=190_{-120}^{+390}$ reported by LVC in \cite{2017ApJ...848L..12A}. In all curves, we found a monotonic decay of the deformability with the increment of the mass}. {In addition, we observe the effects of the quantities $M_{\chi}$ and $k_F^{\dm}$ on the tidal deformability. For an interval of total masses, we note that lower values $\Lambda$ are obtained when larger values of both $M_{\chi}$ and $k_F^{\dm}$ are used. Note that there are curves that are into the interval $\Lambda_{1.4}$ described by LVC in \cite{2017ApJ...848L..12A}.}

\begin{figure}[tb]
 \centering
\includegraphics[trim=0cm 6cm 0cm 6cm, clip, scale=0.41]{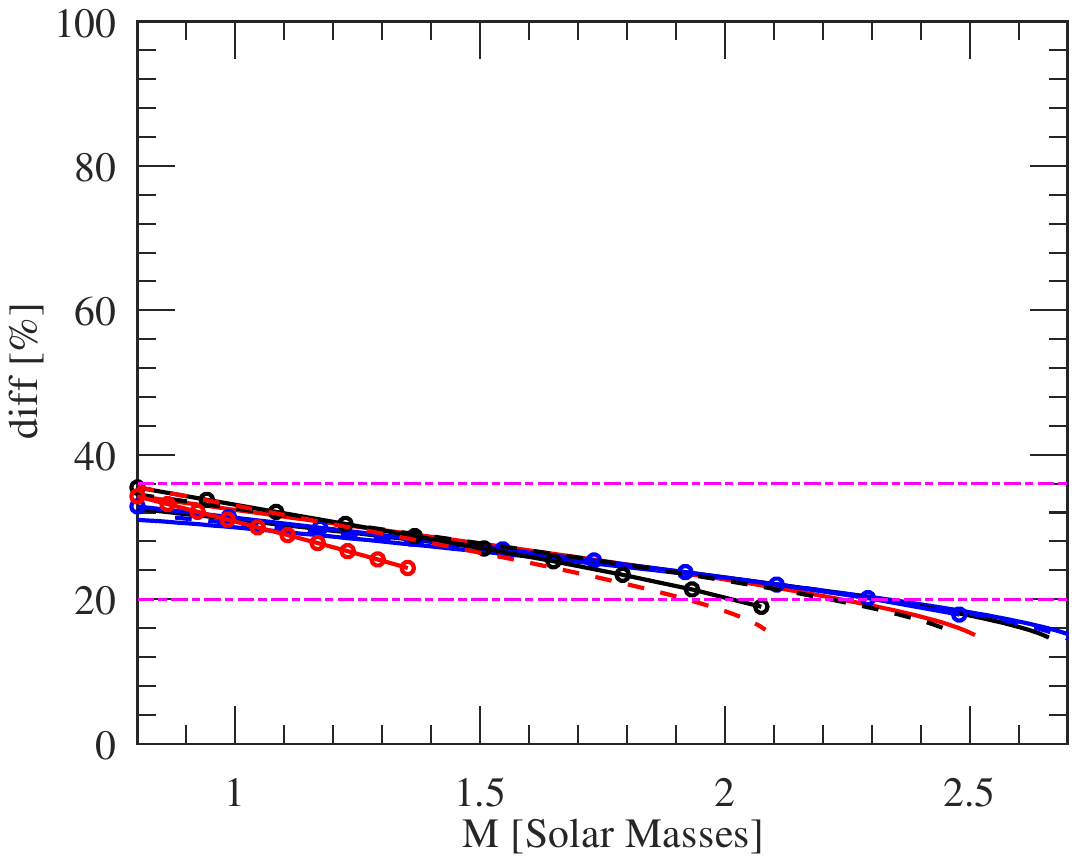} 
\caption{The relative difference between the frequency of the fundamental mode in the full linearized equations and the relativistic Cowling approximation, $diff = 100\times(f_0 - f_{0}^{cow})/f_0$, versus the total mass for all $M_{\chi}$ and $k_F^{\dm}$ considered in Fig. \ref{fig4}.}
\label{fig4}
\end{figure}

Since our objective is to {determine} a relationship between the {mass} stellar {configuration}, tidal deformability, and frequency of the fundamental mode we plot the $f_0$-mode frequency and its respective damping time as a function of the total mass on the top panel of Fig. \ref{fig3} and against the tidal deformability on the bottom panel of Fig. \ref{fig3}, where are employed for different values of $M_{\chi}$ and $k_F^{\dm}$. The results on the bottom panels are compared with the tidal deformability derived from the event GW$170817$, $\Lambda_{1.4}=190^{+390}_{-120}$, communicated in \cite{2017ApJ...848L..12A}. {These plots have} justification in the physics of the binary system because when both stars are orbiting each other, there could exist resonance and, in principle, the fundamental mode of both stars could be excited. {In this way,  in this figure, we see the change in $M_{\chi}$ and $k_F^{\dm}$ produces a systematic shifting in the curves of $f_0(M)$, $\tau(M)$, $f_0(\Lambda)$, and $\tau(\Lambda)$; review also Table \ref{table1} to see the change of the maximum total masses with their respective parameters $R$, $f_0$, and $\tau$ with $M_{\chi}$ and $k_F^{\dm}$ employed in Fig. \ref{fig3}. Moreover, in the range placed by the observation, we see that the $f_0$-mode frequency has a nearly linear behavior. In addition, the frequency of the $f_0$ mode obtained in the full linearized equations of general relativity is compared with the relativistic Cowling approximation in Fig. \ref{fig4}. We found that the $f_0$-mode with the approximation differs by less than about $40\%$, in low-mass stars, and differs by less than about $20\%$, in high-mass stars.}

\begin{figure*}[tb]
 \centering
\includegraphics[trim=0cm 6cm 0cm 6cm, clip, scale=0.269]{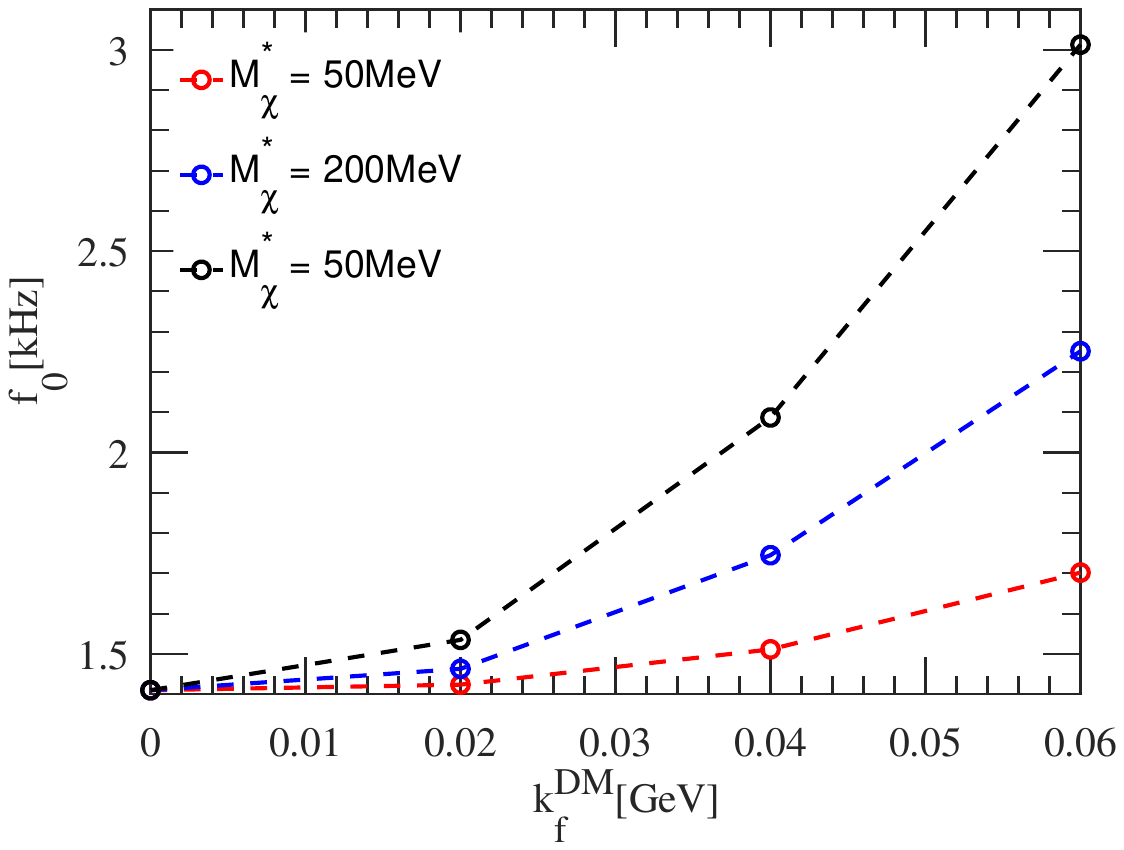} 
\includegraphics[trim=0cm 6cm 0cm 6cm, clip, scale=0.269]{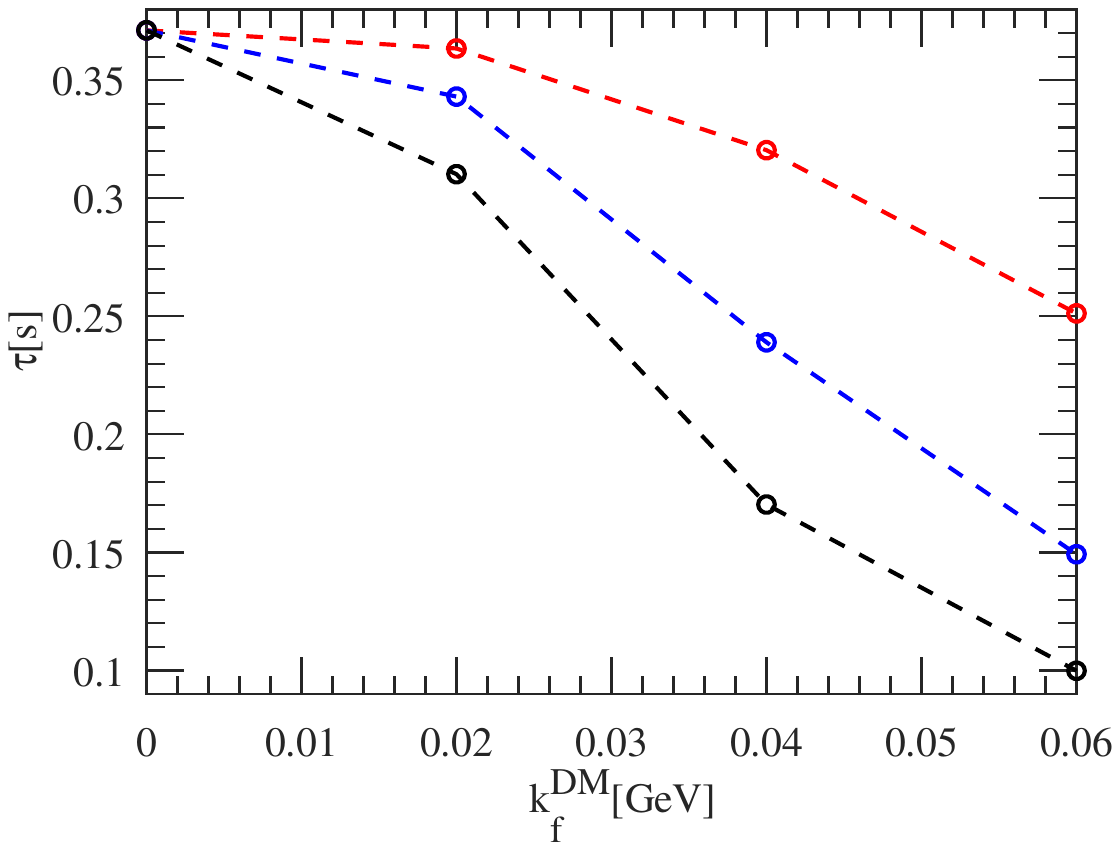} 
\includegraphics[trim=0cm 6cm 0cm 6cm, clip, scale=0.269]{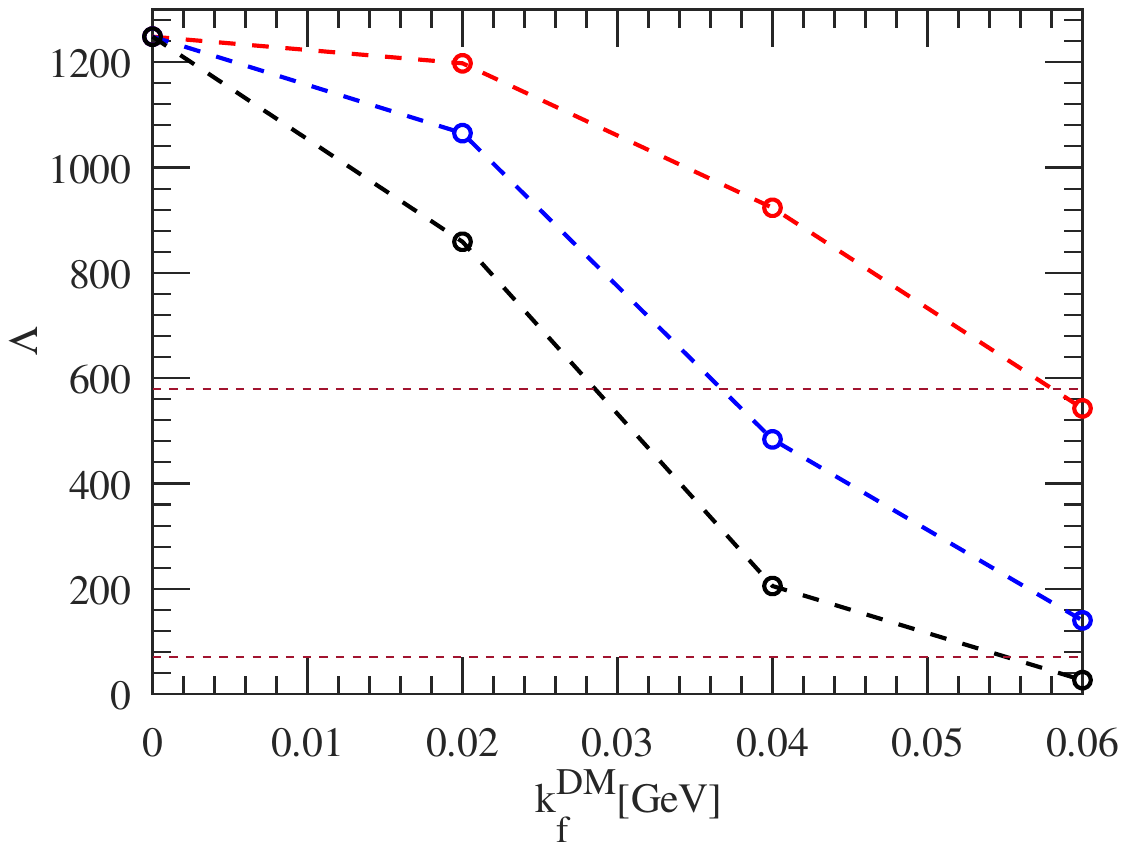} 
\caption{The $f_0$-mode oscillation frequency, damping time parameter, and tidal deformability against the fermi momentum of dark matter are shown respectively on the left, middle, and right panels. The results connected on straight dashed lines on red, blue, and black correspond to $M_{\chi}=50, 200$, and $500$~MeV. All values presented concern the equilibrium configurations with total masses $M=1.4 M_{\odot}$. The horizontal straight line represent $\Lambda_{1.4}=190^{+390}_{-120}$ obtained by LVC in Ref. \cite{2017ApJ...848L..12A}.}
\label{fig5}
\end{figure*}

\begin{figure*}[tb]
 \centering
\includegraphics[trim=0cm 6cm 0cm 6cm, clip, scale=0.269]{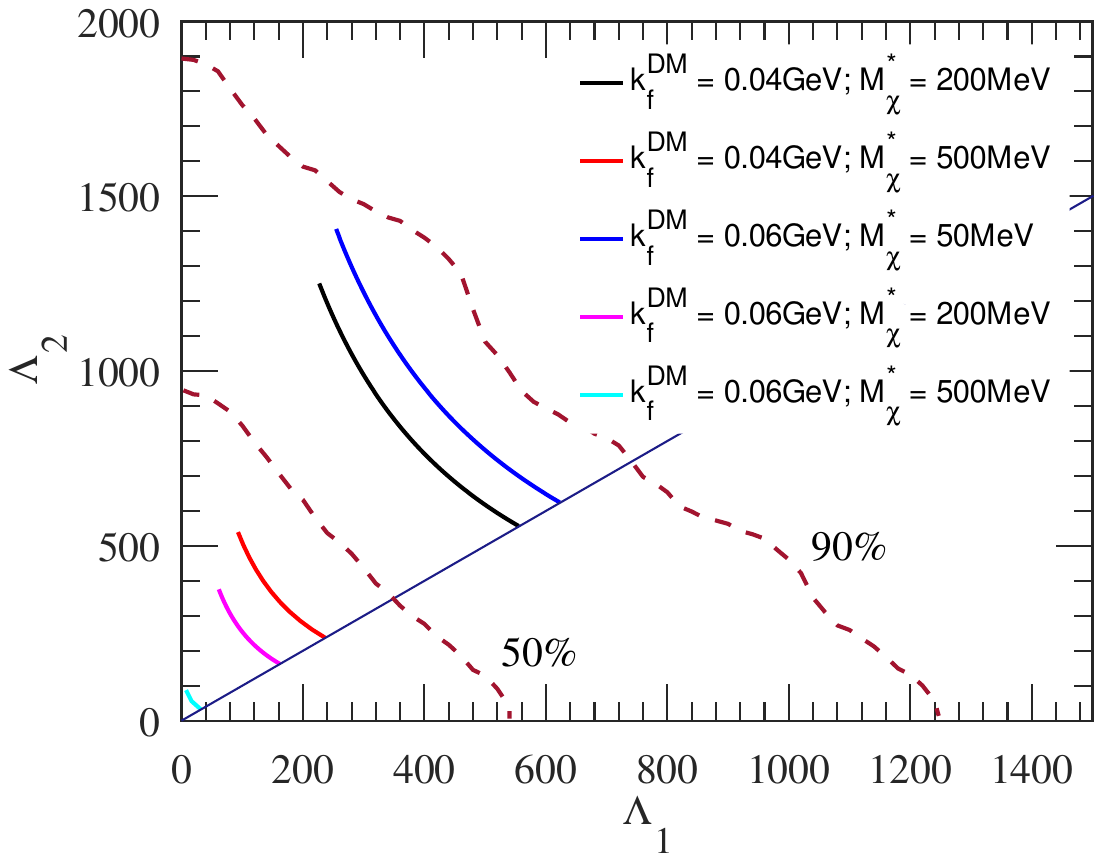} 
\includegraphics[trim=0cm 6cm 0cm 6cm, clip, scale=0.269]{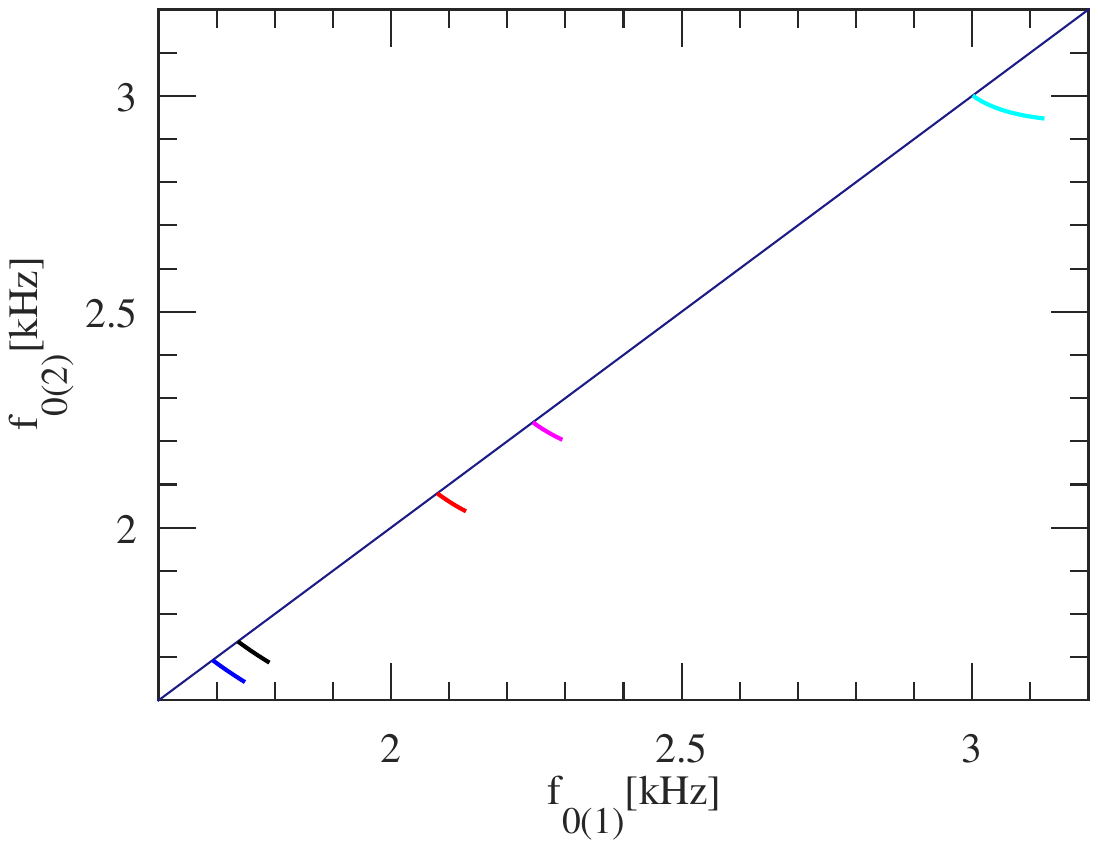} 
\includegraphics[trim=0cm 6cm 0cm 6cm, clip, scale=0.269]{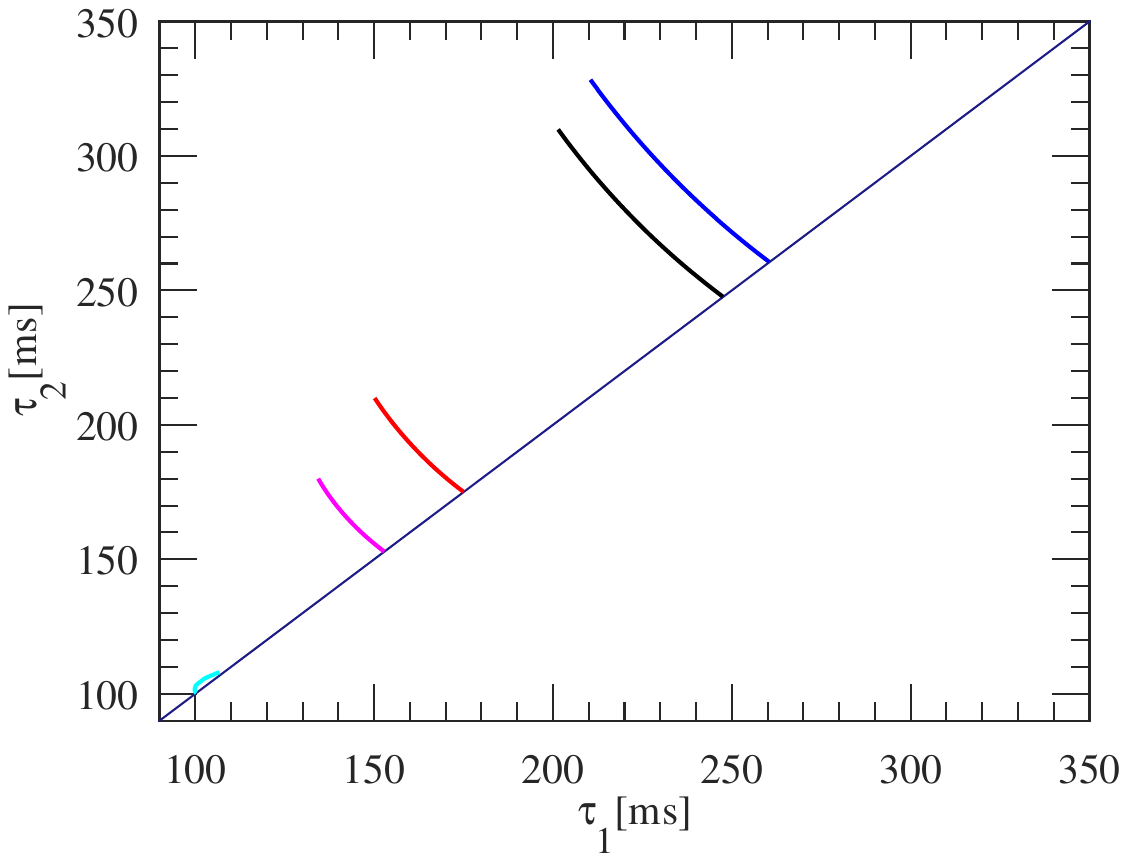} 
\caption{In the Left, middle, and bottom panels are respectively plotted $\Lambda_2\times\Lambda_1$, $f_{0(2)}\times f_{0(1)}$, and $\tau_{2}\times\tau_{1}$ for the components of the event GW$170817$ for different values of $k_F^{\dm}$ and $M_{\chi}$. The wine lines on the left panel depict the LIGO-Virgo confidence curves of $50\%$ and $90\%$ levels \cite{2017PhRvL.119p1101A} and the solid diagonal appearing on the left, middle, and right panels correspond respectively to $\Lambda_1=\Lambda_2$, $f_{0(1)} = f_{0(2)}$, and $\tau_1 = \tau_2$.}
\label{fig6}
\end{figure*}

The $f_0$-mode frequency of oscillations, damping time mode, and tidal deformability as a function of the fermi momentum of dark matter are shown in Fig. \ref{fig5} for three values of dark matter mass $M_{\chi}=50, 200$, and $500$~MeV. The results presented correspond to the equilibrium configuration with total mass $M=1.4M_{\odot}$. The curves $\Lambda\times k_F^{\dm}$ are constrasted with the case $\Lambda_{1.4}=190^{+390}_{-120}$ reported by LVC in Ref. \cite{2017ApJ...848L..12A}.  From the figure, we see that equilibrium solutions with $1.4$ solar masses with larger $f_0$-mode frequency and lower tidal deformability are found when higher values of $k_F^{\dm}$ and $M_{\chi}$ are considered. Comparing the results reported here with those published in \cite{2017ApJ...848L..12A}, we can put limits on the values of $k_F^{\dm}$, and $M_{\chi}$ used. 

\subsection{Tidal deformability, frequency of oscillations, and damping time of fundamental mode for a binary hadronic star system}

In Ref. \cite{2017PhRvL.119p1101A}, by using the data reported by LVC, authors implement some constraints on the dimensionless tidal deformability of a binary system $\Lambda_1$ and $\Lambda_2$, with $\Lambda_1$ representing a star's dimensionless tidal deformability in the system and $\Lambda_2$ depicting the same parameter but of its companion. In this way, in Fig. \ref{fig6}, we present the relation $\Lambda_2\times\Lambda_1$, where the numerical results are obtained by choosing a value of $M_1$ and finding $M_2$ through the chirp mass $M=1.188M_{\odot}$ \cite{2017PhRvL.119p1101A} which is determined by
\begin{equation}
M=\frac{(M1\,M2)^{3/5}}{(M1+M2)^{1/5}}.    
\end{equation} 
In addition, the value of $M_1$ and $M_2$ are respectively in the interval $1.36\,M_{\odot}\leq M_1\leq 1.60\,M_{\odot}$ and $1.17\,M_{\odot}\leq M_2\leq1.36\,M_{\odot}$. In the figure are also placed the credibility levels linked to the GW$170817$ event instituted by LVC in the low-spin prior scenario. We note that the larger values of $k_F^{\dm}$ and lower values of $M_\chi$, and vice versa, allow us to obtain values within the confidence lines extracted from Ref. \cite{2017PhRvL.119p1101A}. In Fig. \ref{fig6} also appears the curves $f_0(2)\times f_0(1)$ and $\tau_2\times\tau_1$ for the same binary system. In these two relations, for all values of $k_F^{\dm}$ and $M_\chi$ considered, we see the dependence of the frequency and the damping time of one star concerning its partner. In the first one, we note that $f_{0(2)}$ decays with the increment of $f_{0(1)}$ and in the second one $\tau_{2}$ grows with the decline of $\tau_{1}$.

\section{Conclusions}\label{conclusion}

In the present article we study the effects of the dark matter mass $M_{\chi}$ and the Fermi momentum of dark matter particles $k_F^{\dm}$ in the fluid pulsation mode, damping time parameter, and tidal deformability of hadronic stars. These studies are developed by the numerical integration of the hydrostatic equilibrium, nonradial oscillation, and tidal deformability equations. The matter within the stars is considered the NL3* equation of state.

About the total mass, it is observed that it changes notoriously with $M_{\chi}$ and $k_F^{\dm}$. From an appropriate combination of $M_{\chi}$ and $k_F^{\dm}$, we can determine that some solutions that are close to the empirical evidence of neutron stars PSR J$0030 + 0451$, PSR J$0740 + 6620$, PSR J$0740+6620$, PSR J$0348+0432$, and PSR J$1614+2230$.

Concerning the fluid pulsation mode, damping time parameter, and tidal deformability, they are considerably affected by $M_{\chi}$ and $k_F^{\dm}$. We found that the $f_0$-mode grows with $M_{\chi}$ and $k_F^{\dm}$ and the damping time and tidal deformability decay with these parameters. We also contrast these results with the observational reported by the LVC from the GW$170817$ event. In this scheme, the numerical results reported are in the range of the observational data aforementioned.

We also calculated the frequencies of the fundamental mode by using the Cowling approximation, as done in the reference \cite{rmfdm10}; where the authors showed that dark matter affects the $f$-mode frequency in the RMF model with and without hyperons.  In our model, it was derived the same effect in the presence of dark matter in hadronic stars. We showed that for larger mass values, the Cowling method provides a good approximation compared to the solution of the complete equation; the difference between these two models is below 20\% for stars with masses above $2 M_\odot$ (see Fig.~\ref{fig4}).

The influence of $M_{\chi}$ and $k_F^{\dm}$ on the tidal deformability, $f$-mode frequency, and damping time parameter for a binary hadronic star system with equal chirp mass as the GW170817 event have been also studied. We found the dependence of $f_{0(1)}$-mode, $\tau_1$, and $\Lambda_1$ of one star with $f_{0(2)}$-mode, $\tau_2$, and $\Lambda_2$ of its partner and how these relations change with the dark matter mass and the fermi momentum of dark matter particles.

\section*{ACKNOWLEDGMENTS}
This work is a part of the project INCT-FNA proc. No. $464898/2014-5$. It is also supported by Conselho Nacional de Desenvolvimento Cient\'ifico e Tecnol\'ogico (CNPq) under Grants No. $401565/2023-8$ (C.H.L., O.L, M.D.), No. $305327/2023-2$ (C.H.L.),  No. $312410/2020-4$ (O.~L.), and No. $308528/2021-2$ (M.~D.). J. D. V. A. thanks Universidad Privada del Norte and Universidad Nacional Mayor de San Marcos for the financial support - RR Nº$\,005753$-$2021$-R$/$UNMSM under the project number B$21131781$. C. V. F. also makes his acknowledgements for the financial support of the productivity program of the  Conselho Nacional de Desenvolvimento Cient\'ifico e Tecnol\'ogico (CNPq), with project number $304569/2022-4$.

\bibliographystyle{apsrev4-2}
\bibliography{references}

\end{document}